\documentclass{iau} 
\usepackage{graphicx}
\usepackage[hyphens]{url}
\usepackage{hyperref}
\hypersetup{
     colorlinks  = true,
     linkcolor   = blue, 
     citecolor   = blue,
     filecolor   = blue,
     menucolor   = blue,
     urlcolor    = blue
}
\usepackage[authoryear]{natbib}

\DeclareGraphicsExtensions{.pdf}

\title[The China crisis]{The China crisis}

\author[Xiaowei Liu]{Xiaowei Liu}

\affiliation{South-Western Institute for Astronomy Research\\
Yunnan University, Chenggong District, Kunming 650500, P.R.\, China
\\email: {\tt x.liu@ynu.edu.cn}}

\pubyear{2018}
\volume{349}  
\jname{Under One Sky -- The IAU Centenary Symposium}
\editors{D. Valls-Gabaud,  C. Sterken\&  J. Hearnshaw   eds.}

\begin{document}

\maketitle

\begin{abstract} The so-called \textit{China crisis}, well documented in \textit{History of the
IAU} by Adriaan Blaauw   and in \textit{Under the Same
Starry Sky: History of the IAU} by Chengqi Fu and Shuhua Ye,   refers to the withdrawal in 1960 of the
People's Republic of China (PRC) from the Union. The crisis stemmed from the
admission by the IAU, amidst strong protest from PRC and some other member
countries, of the Republic of China (ROC) to the Union, creating the so-called
``\textit{Two Chinas}'' -- or ``\textit{One China, one Taiwan}'' problem. The crisis directly led to the
absence of mainland Chinese astronomers from the stage of international
collaborations and exchanges, and was only solved two decades later.
The solution, accepted by all the parties involved, is that China is to have
two adhering organizations, with mainland China astronomers represented by the
Chinese Astronomical Society located in Nanjing (China Nanjing) and China
Taiwan astronomers represented by the Academia Sinica located in Taipei (China
Taipei). The denominations ``\textit{China Nanjing}''  and ``\textit{China Taipei}'' represent the
IAU official resolution and should be used in all IAU events.

The China crisis, probably the most serious one in IAU history, was a
painful lesson in the 100-year development of the Union.  Yet, with its
eventual solution, the Union has emerged stronger, upholding its spirit of
promoting astronomical development through international collaboration of
astronomers from all regions and countries, regardless of the political
systems, religion, ethnicity, gender or level of  astronomical development.

\end{abstract}

\firstsection 
\section{The entry and early involvement of China with the Union}

China has a rich tradition in astronomical observations and studies and boasts the
longest and most comprehensive ancient records of astronomical phenomena dating
back to the Shang dynasty. In its 5000-year continuous civilization, China has
nurtured world-class astronomers including Heng Zhang (78--139 AD), Shoujing Guo
(1231--1316 AD) and Guangqi Xu (1562--1633 AD). 
 
The works of Euclid of Alexandria, Claudius Ptolemy, Nicolas Copernicus and Tycho
Brah\'e were translated, and introduced to and practised by Chinese astronomers in the 
late Ming dynasty, following the arrival of Jesuit priests (Matteo Ricci,
Nicolaus Longobardi, Giacomo Rho, Johann Schreck, Johann Adam Schall von Bell).
In 1644, Johann Adam Schall von Bell (1592--1666 AD) was appointed Director of
the Imperial Observatory, the first time the position was held by a foreigner.

From the \textit{First Opium War} in 1840, China was forced to open up ports under the
weapons of the rising Western powers, and was reduced gradually to a semi-colonial,
semi-feudal country which ushered in a warlord era. In 1912, the last feudal
dynasty Qing fell and the Republic of China (ROC) was founded. The Lugou (Marco
Polo) Bridge Incident on July 7, 1937 marked the beginning of full-out invasion
of China by  Imperial Japan and the Eight-year Chinese War of Resistance which followed, and 
which ended in 1945 along with the World War II. This was followed by a bitter
period of civil war between the ruling Nationalist Party (Kuomingtang -- KMT)
of Jieshi Jiang (Kai-shek Chiang) and the Communist Party of China (CPC) led by
Zedong Mao (Tse-tung Mao). The war ended in 1949 with the founding of the
People's Republic of China (PRC) and the retreat of KMT from mainland China
to the Chinese island of Taiwan. 

The World War II ended with the emergence of two political blocs, the Western
Bloc led by the United States of America (to which the provincial KMT
government in Taipei belonged) and the Eastern Bloc (with which the PRC was 
associated) led by the Soviet Union, which fought between them the Cold War that
only ended decades later in 1992. In 1950, the Korean War broke out between the
Eastern Bloc North Korea (the Democratic People's Republic of Korea -- DPRK)
and the Western Bloc South Korea (the Republic of Korea -- ROK). The war
evolved rapidly into one involving PRC forces on the DPRK side and U.S.  forces
fighting for the ROK. The War ended with the \textit{Korean Armistice Agreement} signed
in 1953.

Several astronomical observatories were built in China by foreign Jesuit
priests in the late nineteenth century, including Shanghai Xujiahui (1873) and
Sheshan (1901) Observatories by the French, Hongkong Observatory (1882) by the British
and Qingdao Observatory (1898) by Germans. In 1922, Qingdao Observatory was
reclaimed by China. Chinese meteorologist and astronomer returning from
the Universit\'e libre de Bruxelles,
Bingran Jiang (1883--1966), became
Director, commencing modern Chinese astronomical observation and research. 

Meanwhile, more Chinese intellectuals studying abroad returned back to China. These included Lu
Gao (1877--1947), inspired by French astronomer Camille Flammarion (1842--1925),
Qingsong Yu (1897--1978) returning from the Lick Observatory, University of
California and Yuzhe Zhang (Yu-Cheh Chang; 1902--1986) from the Yerkes
Observatory, University of Chicago.  In 1926, the Mathematics Department of Sun
Yat-sen University (SYSU) was renamed the Mathematics and Astronomy Department
and began to offer astronomical courses, marking the beginning of astronomical
higher education in China. In 1929, Yun Zhang (1896--1958) who studied astronomy
at the University of Lyon built the SYSU Observatory and became its first Director.
In 1928, the Institute of Astronomy, Academia Sinica was founded.  On October
30, 1922, the Chinese Astronomical Society (CAS) was founded with 47 members at the 
Beijing Ancient Observatory (constructed in 1442 during the Ming Dynasty on the
original site first built by Shoujing Guo in 1279 of the Yuan Dynasty). Lu Gao
became its first President. By 1947, the CAS had 688 individual and 6
institutional members. 

Initiated by Yuanpei Cai (1868--1940), the first Director (1928--1940) of Academia
Sinica and President (1916--1927) of Peking University, Purple Mountain
Observatory (PMO), the best astronomical observatory at the time in the Far East, was inaugurated in
1934. Since its establishment, the CAS took liaison with the Union and
collaboration with the international community was one of its vital tasks. CAS
member Yun Zhang attended in 1925 the 2nd IAU General Assembly (GA) in Cambridge as
an observer, and members Qingsong Yu and Jinyi Zhao participated in 1928 in the 3rd GA in
Leiden. All these individuals paved the way for China joining the IAU. In 1935, at the 5th GA
in Paris, China was formally admitted  to the Union as its 26th National Member, with
the CAS located at PMO in Nanjing as the adhering organization. Four initial
individual members, Qingsong Yu, Lu Gao, Bingran Jiang and a Japanese astronomer
Shinjo Shinzo (1873--1938) were also admitted. At the 6th GA in 1938 in Stockholm, China's
individual members increased to 11. 

In spite of the War, Chinese astronomers did their best to engage with
the Union and paid the arrears in 1947. After the civil war and the
establishment of PRC in 1949,  a delegation of four astronomers from mainland China, 
invited by President Otto Struve (1897--1963), and including Yuzhe Zhang, 
attended in 1955 the Dublin GA. The PRC resumed its legitimate National
Membership in the Union and cleared the arrears. 

\section{The conflict and the withdrawal of China from the Union}

The ``application'' for Union membership from the ROC, submitted in late April
and early May 1958, just two months before the Moscow GA, was not initiated by
astronomers from the island (there were few if any at the time), but was
orchestrated by the US government in order to further isolate the PRC after the
Korean War, at the height of the Cold War.\footnote{The ``Chinese Astronomical
Society'' in Taipei was only established in July 1958, two months after the
application submission.} It came at a time when the US National Committee for
Astronomy (NCA) was about to submit their invitation to the EC to host the 1961
GA in Berkeley, after the 1958 GA  in Moscow. The plot by the US Department of
State, using visas to the US as a threat, was deliberately designed to (one
stone, three birds): 1) Block astronomers from the ``communist'' mainland
China; 2) Promote the status of ``Free China (Taiwan)''; and 3) Create the
so-called ``\textit{Two Chinas}'' or ``\textit{One China, one Taiwan}''
problem. 

The plot was opposed by Leo Goldberg (US NCA), Otto Struve (former IAU
President and still consultant to the EC), and Detlev Wulf Bronk (President,
National Academy of Sciences), worrying that this would tarnish the US
scientific reputation.  In spite of the pressure from the US government, the EC
decided to postpone any decision until after the Moscow GA, in recognition of
``\textit{the serious implications}'' that ``\textit{acceptance of the Taiwan application during
the Moscow meetings might have had: immediate withdrawal of mainland China from
the IAU, and possibly also that of USSR, the host}'' \citep[][p.\,193]{Blaauw}.

Goldberg contacted his representative in the U.S. Congress, George Meader, a
conservative and fair-minded Republican, who presented the case to John Foster Dulles (the US Secretary of State), who
referred it to his science advisor, Wallace Brode. Brode promptly demanded that
Taiwan be invited to the IAU. 

\begin{quotation}``\textit{The fact that Taiwan then had no astronomers and would have to qualify for
IAU membership in the approved way meant nothing to the militant anti-Communist
Brode. Brode wanted Goldberg to go to the 1958 Moscow meeting and submit the
1961 invitation but with the condition that Taiwan be admitted at once. Such a
demand could well wreck the IAU. From Brode's point of view, if the astronomers
would not go along with his orders, so much the worse for them.}'' [Biographical
Memoirs, U.S. National Academy of Sciences, 1997; \cite[cited from][p.\,154]{FY09}].\end{quotation}

Unfortunately, after the Moscow GA, the attitude of the EC made a U-turn.
In spite of opposition by the IAU Vice-Presidents from the USSR and Czechoslovakia,
who stated that admission should be judged solely on scientific grounds, that
the astronomical activity in this applicant's country was too low, and that the
admission of Taiwan might risk the withdrawal of mainland China, the EC pushed
through the admission by the ballot, with five votes for and two against.

Yuzhe Zhang, President of the CAS in Nanjing, in his letter to J.H. Oort,
serving President of the IAU, expressed surprise and indignation: 

\begin{quotation}``\textit{\ldots Taiwan
is an inseparable part of Chinese territory, it is a province of China\ldots
Should the report be authentic, I, on behalf of the Astronomical Society of the
People's Republic China, hereby lodge our strong protest with you and insist
that the Executive Council of IAU rescind the illegal decision\ldots Otherwise,
the Astronomical Society\ldots  will resolutely and definitely withdraw from the
IAU''} \citep[][p.\,193]{Blaauw}.\end{quotation}

IAU \textit{Information Bulletin} (IB) No.\,2 of November 1959 announced the adherence of
Taiwan as a member of the IAU. The withdrawal of mainland China was announced in
IB No.\,3 of May 1960. The rapid announcements underscored the acceptance as a
``\textit{fait accompli}''.

The EC further brushed aside letters of protest from the Polish and Bulgarian
Academies sent in March 1960, as well as concerns raised by Vice-President O.
Heckmann during the 1961 GA. Also during the Assembly, President Oort, in an
unusual move, before the vote, asked representatives to vote against a combined
motion submitted by the USSR and Czechoslovakia Academies of Sciences
requesting the decision of the EC to admit Taiwan be revoked.

The decision of the EC, succumbed to the political pressure from the US
government, led to the two-decade absence of astronomers of mainland China from
the international stage of astronomical collaboration and exchange. Squeezed
between the demands of the then two superpowers, the United States of America
and the Soviet Union, the IAU survived, but unfortunately was clearly damaged. 

\section{The return of China}

In the following two decades, in spite of the growing dissatisfaction and
concern, restrained by the then prevailing political environment, both
international and domestic, little progress was made on the restoration of the
legitimate position of China in the IAU until the 1979 Montreal Assembly. Several events before then eased the way forwards from the prevailing deadlock:
\begin{enumerate}
\item[1.] In 1971, UN Resolution No.\,2758 restored the membership of the PRC and expelled
the ROC from the UN; \item[2.] In 1972, Richard Nixon visited China. The joint communiqu\'e
recognized Taiwan as a part of China, thereby ushering in a new era of Sino-American
relations; \item[3.] In 1976, the ten-year Cultural Revolution in mainland China
ended; and \item[4.] In 1978, China started the economic reforms and opening up. 
\end{enumerate}
All these paved the political way for China rejoining the Union.

The key was to find ``\textit{a way acceptable to the Chinese of reinstating their
membership in the IAU without expelling Taiwan, an action that would violate
the statutes of the Union}'' \citep[see p.20]{Goldberg}.


Initiated by President Adriaan Blaauw, under the invitation of the EC, a
six-member delegation from mainland China, including Yuzhe Zhang, Shuhua Ye
(Vice President of the IAU 1988--1994) and Zhaohua Yi, and a single-member
delegation from Taiwan represented by C.S. Shen, President of the Taiwan NCA,
arrived in Montreal to discuss the matter prior to the GA on August 13--24 1979,
and to explore the possibility of ``dual membership'', that, on the one hand,
expresses the indivisibility of China (emphasized and agreed by both parties
across the Straight), and, on the other hand, reinstates the membership of PRC
without blocking China Taiwan's further adherence.

The negotiations resulted in a proposal communicated to the GA at its closing
session, that was presented in the form of an exchange of letters dated
Montreal, August 22, 1979, between the President of the CAS, Yuzhe Zhang, and
the outgoing President of the IAU, Adriaan Blaauw. The two letters were reproduced in
full in IAU \textit{Transactions} Vol. XVII \citep[pp 48--50]{Bappu}.
The only problem left was the name of the adhering organization in Taiwan.

\begin{figure}[t]
\begin{center}
\includegraphics[width=0.8\textwidth]{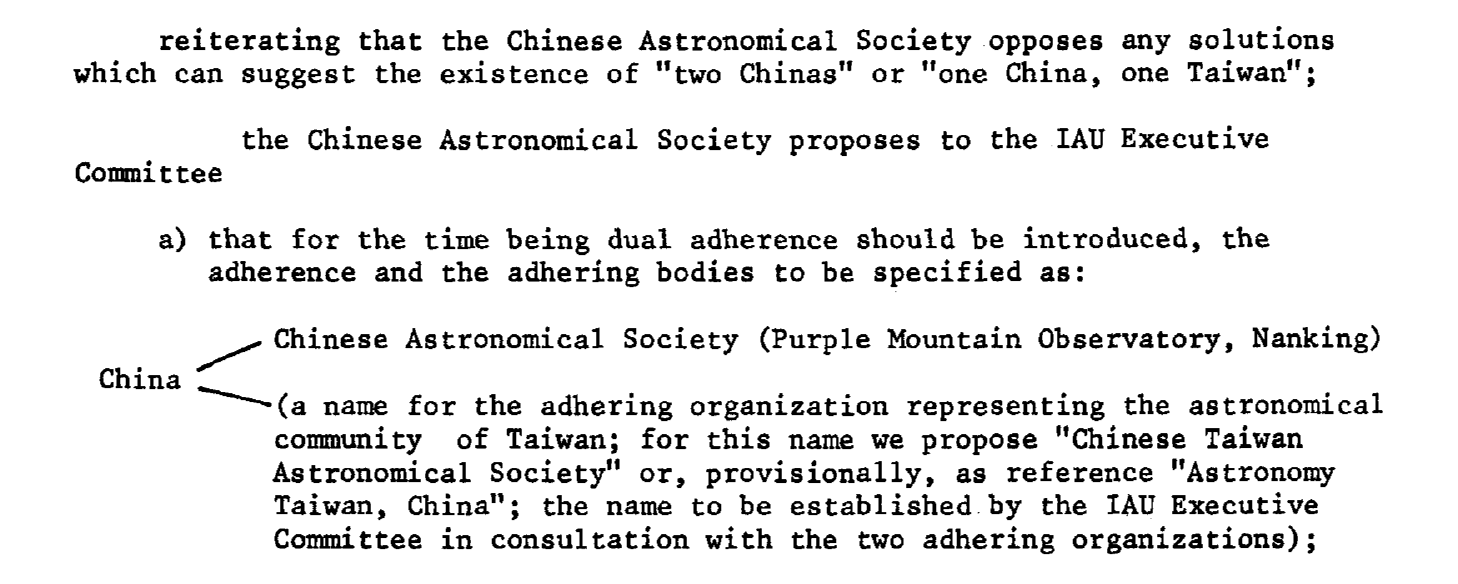} 
 \caption{Parts of the letter from Yuzhe Zhang, President of the Chinese 
Astronomical Society (1943--1982), to Adriaan Blaauw, President of the IAU \citep[pp. 525--527]{ChangBlaauw}}
   \label{fig1}
\end{center}
\end{figure}

The problem was shortly solved in 1980. The arrangements were ratified at the
1982 Patras GA XVIII.  The solution, accepted by all the parties involved, is
that China is to have two adhering organizations, with  mainland Chinese
astronomers represented by the Chinese Astronomical Society located in Nanjing
(\textit{China Nanjing}; since 1935) and \textit{China Taiwan} astronomers represented by the
Academia Sinica located in Taipei (China Taipei; since 1959). The
denominations, ``\textit{China Nanjing}'' and ``\textit{China Taipei}'' represent the IAU
official resolution and should be used in all IAU events. At the time of this
writing (August, 2018), ``\textit{China Nanjing}'' and ``\textit{China Taipei}'' have
respectively 666 and 74 IAU individual
members.\footnote{\href{https://www.iau.org/administration/membership/national/}{www.iau.org/administration/membership/national}}

\begin{quotation}``\textit{The Union had succeeded in overcoming political schism and in restoring
harmony again among its membership for its prime purpose: the unhampered
pursuit of scientific research and intercourse}'' \citep[][p.\,204]{Blaauw}.
\end{quotation}

Shouguan Wang (President of the CAS 1985--1989), on behalf of the CAS, gave a
warm speech after the ratification of the membership of China by the GA \citep[p. 26]{Wang}:
\begin{quotation}\textit{Dear friends \& colleagues,} 

\textit{The Chinese Astronomical Society celebrates its 60th anniversary this year. Its
reunion with this international community today is an event that is highly
appreciated by all its 900 members. I and my colleagues here are very glad to
have this opportunity of speaking on behalf of our Society and its members to
express our most cordial greetings and most sincere thanks to you all. Thank
you!}   \end{quotation}

Zden$\check{\rm e}$k Kopal, writing on \textit{The IAU --  the first 60 years.
Reminiscences and Reflections}, published in \textit{Astrocosmos}, the newspaper of the
Patras GA, wrote, \begin{quotation}``\textit{\ldots the Union has really never been free of political
interference from many directions ever since. Perhaps the most conspicuous
example of such an interference in recent years was the technical expulsion of
the (People's) Republic of China, which was eased out of our midst in 1955
(1960) by the United States (during the enlightened era of John Foster Dulles),
in collaboration with certain astronomers from Western Europe. Only God knows
what good should have come to the science of astronomy and to the International
Astronomical Union from severing (albeit temporarily) its official ties with
the most populous nation of the Earth; but such acts did happen, and will
continue to happen as long as the present structure of the IAU remains
unchanged}''  \citep{Kopal}. \end{quotation}

The resolution of the China crisis, combined with the continued improvement of
relation across the Taiwan Straight, has benefited astronomers on both sides,
stimulated and facilitated collaborations amongst them as well as with the
international community at large. Astronomy in both mainland China and in the
Island has entered an era of rapid development.  Four decades after the crisis,
China is poised to make major contributions to the world astronomy research and
education development.

In 2012, the IAU GA XXVIII was held in Beijing, being the first time in China in the
Union's nearly hundred-year history. In his opening ceremony speech, Vice-President 
Jinping Xi remarked: 
\begin{quotation}``\textit{The development of science and technology
requires extensive international cooperation. Science and technology have no
nationality! The vast expanse of space is the common home of all humankind; to
explore this vast universe is the common goal of all humankind; astronomy in
fast development is the shared fortune of all humankind}''  \citep{Xi}. \end{quotation} 


\begin{thebibliography}{}

\bibitem[Bappu\,(1979)]{Bappu}Bappu, M.K.V. 1979,  Report of the XVIIth General
Assembly of the IAU,  {\it Trans. of the IAU}, 17B, 15, publ.  by Association
of Univ. for Research in Astronomy

\bibitem[Blaauw\,(1994)]{Blaauw}Blaauw, A. 1994, \textit{History of the IAU:
The Birth and First Half-Century of the International Astronomical Union}
(Dordrecht: Kluwer), 296 pp.

\bibitem[Fu and Ye\,(2009)]{FY09}Fu, Chengqi  and Ye, Shuhua  2009,
\textit{Under the Same Starry Sky. History of the IAU} (Shanghai: Jiaotong
University Press 

\bibitem[Goldberg\,(1977)]{Goldberg}Goldberg, L. 1977, {\it Trans. of the IAU},
16B, 15, publ.  by Association of Univ. for Research in Astronomy

\bibitem[Kopal\,(1982)]{Kopal}Kopal, Z. 1982, {\it Astrocosmos} (Newspaper of
the IAU GA in Patras, Greece) 2, 5

\bibitem[Wang\,(1983)]{Wang}Wang, S.G. 1983, {\it Trans. of the IAU}, 18B, 15,
publ.  by Association of Univ. for Research in Astronomy

\bibitem[Xi\,(2012)]{Xi}Xi, J.P. 2012, \textit{Inquiries of Heaven}, (Newspaper
of the IAU GA in Beijing, Wednesday, August 22, 2012) 3, 1

\bibitem[Zhang \& Blauuw\, (1979)]{ChangBlaauw}Zhang, Y.Z. (Chang,
Y.C.) and Blaauw, A. 1979, Apprendix IV, Letters exchanged between the
President of the IAU and the President of the Chinese Astronomical Society at
Nanking, {\it Trans. of the IAU}, 17B, 525--527, publ.  by Association of Univ.
for Research in Astronomy

\end{thebibliography}
\end{document}